# A Trustworthy Industrial Fault Diagnosis Architecture Integrating Probabilistic Models and Large Language Models

WuYue

Xi'an Jiaotong Uinversity

wuyue0619@stu.xjtu.edu.cn

**Abstract:** Addressing the core problem of insufficient trustworthiness in industrial fault diagnosis, stemming from the limitations of existing methods—both traditional and deep learning-based—in terms of interpretability, generalization, and uncertainty quantification, this paper proposes a trustworthy industrial fault diagnosis architecture, the Hierarchical Cognitive Arbitration Architecture (HCAA), which integrates probabilistic models with Large Language Models (LLMs). The architecture conducts a preliminary analysis via a diagnostic engine based on a Bayesian network and features an LLM-driven cognitive arbitration module with multimodal input capabilities. This module performs expert-level arbitration on the initial diagnosis by analyzing structured features and diagnostic charts, holding the priority to make the final decision upon detecting conflicts. To ensure the reliability of the system's output, the architecture integrates a confidence calibration module based on Temperature Scaling and a risk assessment module, which objectively quantify system trustworthiness using metrics like Expected Calibration Error (ECE). Experimental results on a dataset containing multiple fault types demonstrate that the proposed framework improves diagnostic accuracy by over 28 percentage points compared to baseline models, while the post-calibration ECE is reduced by more than 75%. Case studies confirm that the HCAA effectively corrects misjudgments from traditional models caused by complex feature patterns or knowledge gaps, providing a novel and practical engineering solution for building high-trust, explainable AI diagnostic systems for industrial applications.

1. Introduction

With the deep development of Industry 4.0 and smart manufacturing concepts, modern industrial systems are evolving towards high levels of automation and intelligence. In this process, the reliability and safety of equipment have become key factors determining production efficiency and operational costs. Prognostics and Health Management (PHM), as a core technology, plays an indispensable role in improving equipment reliability, reducing unplanned downtime, and optimizing maintenance costs by monitoring equipment status in real-time, diagnosing potential faults, and predicting remaining useful life [1], [2]. A successful PHM system can not only avoid catastrophic failures but also transform the traditional reactive maintenance model into a proactive, predictive one, thereby bringing enormous economic and safety benefits [3].

Although PHM technology has made significant progress, it still faces a series of severe challenges in practical engineering applications. Traditional methods based on physical models or shallow machine learning heavily rely on expert experience for feature extraction and rule-making, which not only leads to long development cycles and high costs but also suffers from poor generalization, making it difficult to adapt to varying working conditions and diverse equipment [4]. In recent years, data-driven methods, particularly deep learning, have shown great potential in the field of fault diagnosis [5], capable of automatically learning fault features from massive data. However, these end-to-end deep learning models are typically "black boxes," whose internal decision-making logic is difficult to interpret, leading engineers to lack trust in their diagnostic results [6]. Furthermore, these models are highly sensitive to the distribution of training data; when faced with new working conditions or equipment different from the training set, their performance often degrades sharply, indicating severely insufficient cross-condition generalization ability. More critically, these models universally lack effective quantification of their predictive uncertainty, making it impossible for users to assess the reliability of a diagnostic result, which is unacceptable in industrially safety-critical fields, as an incorrect diagnostic decision could lead to catastrophic consequences [7]. Therefore, building trustworthy AI diagnostic systems— that is, providing high accuracy while possessing interpretability, strong generalization capability, and reliability quantification—has become a core problem to be solved in the current field of industrial intelligence [8], with its importance even surpassing mere accuracy improvement.

The advent of Large Language Models (LLMs), especially their emergent capabilities in reasoning, multimodal understanding, and knowledge integration, has brought unprecedented opportunities to solve the aforementioned bottlenecks in the PHM field [9], [10]. By pre-training on massive text data, LLMs can master deep domain knowledge and apply complex logical reasoning chains to analyze and judge problems [10]. Recently, Tao et al. pioneeringly proposed the concept of PHM Large Models and systematically elaborated on its concepts, paradigms, and challenges, pointing the way for the deep integration of AI technology and PHM [1]. Within this framework, the parallel paradigm is particularly noteworthy, as it advocates for the collaboration of traditional PHM models with large models to combine the former's expertise in signal processing and the latter's superiority in knowledge reasoning. Compared to directly building an end-to-end PHM large model, the parallel paradigm is more engineering-feasible at the current stage because it can fully leverage existing mature signal processing and diagnostic models, allowing LLMs to focus on their strongest suit—knowledge integration and logical reasoning—thereby achieving complementary advantages and rapid implementation. However, this theoretical framework also raises new, more specific research questions: In engineering practice, how should this parallel collaboration be specifically implemented? When the analysis results of traditional PHM models conflict with the reasoning conclusions of large models, who should the system trust? More importantly, how can we ensure that the LLM's output is not based on hallucinations [11] from its training data but is a genuine and reliable diagnostic basis? These issues are directly related to whether LLMs can be truly implemented in highly trustworthy industrial scenarios.

To address these challenges, this paper proposes a Hybrid Cognitive Arbitration Architecture based on Large Language Models for trustworthy industrial fault diagnosis. The main contributions of this work are as follows: First, to the best of our knowledge, it is the first time that the concept of "conflict arbitration" has been systematically applied to the field of industrial fault diagnosis, designing a novel HCAA framework that synergistically combines a probabilistic model-based diagnostic engine with a Large Language Model. Second, we designed and implemented an LLM-driven cognitive arbitration module based on multimodal input. This module not only analyzes structured data but also performs visual analysis by understanding diagnostic charts (such as spectrograms and order plots) and is endowed with the priority to overturn the diagnostic results of traditional models. Third, we integrated a confidence calibration and risk quantification module, which calibrates the diagnostic confidence through techniques like Temperature Scaling and quantifies the system's reliability using metrics like the Risk-Coverage curve, providing an objective, quantifiable standard for the system's trustworthiness. Finally, through comprehensive experiments and typical case studies on a simulated dataset containing various typical faults, we validated the effectiveness of HCAA in improving diagnostic accuracy and trustworthiness. This study aims to provide a novel and practical engineering solution for high-trust, explainable AI diagnostic systems in industrial environments.

2. Related Work

2.1 Traditional Industrial Fault Diagnosis Methods

The development of industrial fault diagnosis technology has evolved from simple signal processing to complex intelligent models. Traditional methods can be broadly classified into three categories: signal processing-based methods, model-based methods, and data-driven methods.

Signal processing-based methods are a classic means of fault diagnosis, primarily involving time-domain, frequency-domain, and time-frequency-domain analysis of sensor signals such as vibration, acoustics, and temperature to extract features related to faults. Common techniques include Fast Fourier Transform (FFT) to identify harmonic components in the spectrum [12], Wavelet Transform to analyze the time-frequency characteristics of non-stationary signals [13], and Envelope Analysis to extract shock characteristics of rolling bearing and gear faults [14]. These methods are intuitive and physically meaningful, but they heavily rely on expert experience for setting thresholds and interpreting results, exhibiting low automation and difficulty in adapting to complex and variable working conditions.

Model-based methods rely on establishing mathematical models based on the physical mechanisms of the system or historical statistical data. Physical models simulate the responses under different fault states by establishing dynamic equations of the equipment and comparing them with actual observations for diagnosis [15]. Bayesian Networks, as a type of probabilistic graphical model [16], fuse expert knowledge and observational data to calculate the posterior probability of various faults, capable of handling uncertainty and performing inference. These methods have strong interpretability, but their model accuracy heavily depends on a clear understanding of the system's physical mechanisms. For modern industrial systems with complex structures and unclear mechanisms, establishing precise physical models is often extremely challenging.

Data-driven methods, especially models represented by deep learning [17], have achieved great success in recent years. Shallow machine learning models like Support Vector Machines (SVM) and Random Forests, as well as deep learning models like Convolutional Neural Networks (CNN) and Recurrent Neural Networks (RNN), have been widely used for automatic fault classification and localization [18], [19]. These models can directly learn complex fault patterns from data, avoiding the dependence on expert knowledge and physical models. However, deep learning models are often regarded as "black boxes," with opaque internal decision-making logic, making results difficult to interpret and trust [19]. Furthermore, these models are highly sensitive to the distribution of training data; their performance significantly degrades when applied to new working conditions or equipment different from the training set, exhibiting poor cross-condition generalization ability [20], [21]. More importantly, these models universally lack effective quantification of their predictive uncertainty, making it impossible for users to assess the reliability of a diagnostic result, which is unacceptable in safety-critical domains.

2.2 Applications of Large Language Models in the Industrial Sector

In recent years, Large Language Models based on the Transformer architecture [13] have achieved revolutionary breakthroughs in the field of natural language processing. By pre-training on massive text data, LLMs such as the

GPT series [22] and BERT [23] have demonstrated powerful contextual understanding, knowledge integration, and logical reasoning capabilities [10]. These emergent abilities have enabled them to move beyond traditional text tasks and begin to expand into multimodal understanding and cross-domain applications.

In the industrial sector, exploratory applications of LLMs have already emerged, showing immense potential. In expert systems, LLMs can be trained to act as "virtual experts" with deep domain knowledge, providing engineers with troubleshooting suggestions, maintenance plan generation, and technical support through natural language interaction. In report generation, LLMs can automatically integrate diagnostic results, sensor data, and maintenance records into clear, well-structured, and fluently written analysis reports, greatly improving documentation efficiency [24]. Additionally, some research has begun to explore using the code generation capabilities of LLMs to assist in developing or optimizing diagnostic algorithms [25], as well as enhancing information retrieval efficiency through natural language Q&A in interactive maintenance assistants [26].

The core advantage of LLMs lies in their superior ability to process unstructured knowledge and perform logical reasoning. Unlike traditional AI systems that rely on structured databases and fixed rules, LLMs can learn and refine knowledge from a vast amount of technical documents, repair manuals, and case reports, and reason and judge in a manner close to human experts. This capability enables them to effectively address the challenges of scattered knowledge and rapid iteration in industrial environments, integrating fragmented knowledge into coherent diagnostic evidence. Their powerful multimodal understanding capabilities [27], [28], especially the ability to analyze images and charts, provide new avenues for processing diagnostic visualization data such as vibration spectrograms and order plots.

## 2.3 Trustworthy AI and Model Calibration

As AI technology is increasingly applied in safety-critical domains, "Trustworthy AI" [29], [30] has become a research hotspot. Trustworthy AI requires not only high model accuracy but also emphasizes the interpretability, reliability, and robustness of the decision-making process [31]. Explainable AI (XAI) aims to open the "black box," making the model's decision-making logic understandable and auditable by humans. However, achieving complete interpretability for complex deep learning models remains a huge challenge.

Model calibration is an important technical means to enhance the trustworthiness of AI systems. A well-calibrated model's output confidence should truly reflect the correct probability of its prediction. However, in practice, many modern deep learning models suffer from "over-confidence" or "poor confidence" problems [19]. To address this, researchers have proposed various calibration methods, such as Platt Scaling [32], Temperature Scaling, and Isotonic Regression [33]. These methods typically train an additional calibration model on a validation set to map the original model's output confidence to a calibrated probability that is closer to the actual accuracy rate.

To evaluate calibration effectiveness, Expected Calibration Error (ECE) [34] has become a widely accepted metric. ECE quantifies the model's calibration degree by grouping predictions by confidence intervals and calculating the weighted average difference between accuracy and average confidence within each interval. A lower ECE value indicates that the model's confidence is more reliable. In high-risk scenarios like industrial fault diagnosis, a calibrated model can provide a more reliable basis for risk assessment in operational decisions. For example, the system can set a rule to trigger a shutdown check only when the diagnostic confidence is higher than a calibrated threshold, thereby avoiding missed detections while minimizing unnecessary downtime.

## 2.4 Relationship Between This Work and Existing Work

Tao et al. [1] systematically constructed the concept of PHM-LM and proposed three major paradigms, providing a grand blueprint for the deep integration of AI and PHM. Inspired by this, we not only built a parallel system that integrates a traditional Bayesian network diagnostic engine with a Large Language Model but, more importantly, we directly confront and attempt to solve several core problems implicit in this paradigm: How should this parallel collaboration be specifically implemented? When the analysis results of traditional PHM models conflict with the reasoning conclusions of large models, who should the system trust? More importantly, how can we ensure that the LLM's output is not based on hallucinations [11] from its training data but is a genuine and reliable diagnostic basis?

Compared to existing exploratory work applying LLMs to specific domains [24], [26], [27], our work is more focused on solving the core pain point of "trustworthiness" in industrial diagnostics. Compared to research on single fault diagnosis algorithms or uncertainty analysis [35], this paper provides a systematic hybrid cognitive arbitration architecture (HCAA) that integrates cutting-edge AI technology with traditional engineering methods, reliability quantification and assessment, and post-hoc auditability.

## 3. Hybrid Cognitive Arbitration Architecture

### 3.1 Overall Framework

The overall workflow of the HCAA system is illustrated in Figure 1. The data flow and interactions between modules can be abstracted into the following mathematical process.

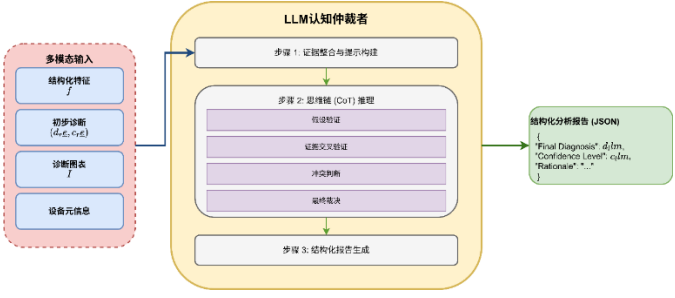

**Figure 1 HCAA System Architecture Diagram**

Given a raw vibration time-series signal $x(t)$ and a set of diagnostic charts $\mathcal{I}$, the diagnostic process of HCAA can be formally described as a series of mappings and decisions:

$$f = \mathcal{E}(x(t)), \quad (d_{\text{rule}}, c_{\text{rule}}) = D_R(f), \quad (d_{\text{llm}}, c_{\text{llm}}) = D_A(f, \mathcal{I}, d_{\text{rule}}, c_{\text{rule}}) \quad (1)$$

$$(d_{\text{arb}}, c_{\text{arb}}) = \mathcal{A}(d_{\text{rule}}, c_{\text{rule}}; d_{\text{llm}}, c_{\text{llm}}), \quad (d_{\text{final}}, c_{\text{cal}}) = C(d_{\text{arb}}, c_{\text{arb}}) \quad (2)$$

Where: $\mathcal{E}(\cdot)$ is the feature extraction module, mapping the raw signal $x(t)$ to a multi-dimensional feature vector $f$. $D_R(\cdot)$ is the diagnostic engine based on a probabilistic model, which outputs a preliminary diagnosis $d_{\text{rule}}$ and its confidence $c_{\text{rule}}$ based on the feature vector $f$. $D_A(\cdot)$ is the LLM-driven cognitive arbitration module, which receives $f$, $\mathcal{I}$, $d_{\text{rule}}$, and $c_{\text{rule}}$, and outputs an arbitration result $d_{\text{arb}}$ and its confidence $c_{\text{arb}}$. $\mathcal{A}(\cdot)$ is the arbitration strategy function, used to handle conflicts between the rule engine and the LLM. $C(\cdot)$ is the confidence calibration module, which receives the final diagnosis $d_{\text{arb}}$ (or $d_{\text{rule}}$, if no conflict) and its original confidence, and outputs a calibrated confidence $c_{\text{cal}}$.

The final output of the system is a trustworthy diagnostic report containing the diagnosis type, calibrated confidence, and a detailed analysis report. This layered design ensures that the system's diagnostic capabilities have both data-driven objectivity and expert knowledge-driven deep insight, while guaranteeing the reliability of its output through the calibration mechanism.

*3.2 Probabilistic Model-Based Diagnostic Engine*

The rule-based diagnostic engine is the baseline of the HCAA system, responsible for performing a rapid, preliminary diagnosis based on historical data and expert rules. It consists of a feature extraction module and a Bayesian network diagnostic module.

3.2.1 Feature Extraction Module

To comprehensively characterize the equipment status from the vibration signal $x(t)$, we extract four categories of key features, forming a combined feature vector $f = [f_{\text{time}}, f_{\text{freq}}, f_{\text{order}}, f_{\text{env}}]^T$. The signal processing details are as follows: Order analysis relies on the shaft frequency $f_s$ obtained from a tachometer, and anti-jitter filtering is applied before synchronous resampling to ensure the accuracy of spectral analysis. The FFT calculation uses a Hanning window with the number of points set to 4096 to ensure sufficient frequency resolution. Before envelope analysis, a band-pass filter (4th-order Butterworth) is used to select the frequency band containing the resonance peak, followed by Hilbert transform demodulation.

**Time-domain features ($f_{\text{time}}$):** These features directly reflect the statistical and impulsive characteristics of the signal.

Root Mean Square (RMS): $\text{RMS}(x) = \sqrt{\frac{1}{N}\sum_{i=1}^{N} x_i^2}$

Crest Factor: $\text{CF}(x) = \frac{\max(|x|)}{\text{RMS}(x)}$

Kurtosis: $\text{Kurtosis}(x) = \frac{E[(x-\mu)^4]}{\sigma^4}$
Where $N$ is the number of signal sampling points, and $\mu$ and $\sigma$ are the mean and standard deviation of the signal, respectively. High kurtosis and crest factor are often important indicators of bearing faults or mechanical shocks.

**Frequency-domain features ($f_{\text{freq}}$):** The signal is converted to the frequency domain via Fourier transform to identify frequency components related to rotational speed.

Dominant Frequency: $f_{\text{dom}} = \underset{f}{\operatorname{argmax}}(|\mathcal{F}\{x(t)\}(f)|)$

Spectral Centroid: $f_c = \frac{\sum_k f_k |X_k|}{\sum_k |X_k|}$
Where $X_k$ is the Fourier transform coefficient of the signal at frequency $f_k$. Abnormalities such as misalignment and gear meshing frequencies directly affect these indicators.

**Order analysis features ($f_{\text{order}}$):** Frequency-domain features are normalized relative to the shaft frequency $f_s$ to analyze harmonic patterns unique to rotating machinery.

1X, 2X Amplitudes: $A_{1X}, A_{2X}$ are the spectral amplitudes at $f_s$ and $2f_s$, respectively. A high ratio of $A_{2X}/A_{1X}$ is often strong evidence of parallel or angular misalignment.

Harmonic Count: $N_h = |\{k | A_{k \cdot f_s} > \tau, k \in \mathbb{N}^+\}|$, where $\tau$ is a preset amplitude threshold. Mechanical looseness usually causes rich harmonic components.

**Envelope analysis features ($f_{\text{env}}$):** The signal envelope is obtained via the Hilbert transform to detect high-frequency shock events, particularly suitable for rolling bearing and gear faults.

Envelope Kurtosis: $\text{EnvKurt}(x) = \text{Kurtosis}(|\mathcal{H}\{x(t)\}|)$, where $\mathcal{H}\{\cdot\}$ is the Hilbert transform.

Envelope Peak Frequency: $f_{\text{env\_peak}} = \underset{f}{\text{argmax}}(|\mathcal{F}\{|\mathcal{H}\{x(t)\}|\}(f)|)$.

### 3.2.2 Bayesian Network Diagnostic Module

We employ a Naive Bayes classifier as the baseline diagnostic model because it not only provides a diagnosis but also an interpretable posterior probability.

**Model Definition:**
Let the set of fault types be $\mathcal{F} = \{f_1, f_2, \ldots, f_M\}$, and each dimension $f_i$ of the feature vector $f$ is discretized into several intervals $I_{i,j}$.

**Parameter Learning:**

**Prior Probability:** $P(F = f_m)$ is set based on historical fault statistics.

**Likelihood Function:** To address the strong feature independence assumption in Naive Bayes and the roughness of discretization, we use a class-conditional Gaussian assumption for continuous features (GaussianNB), and apply Laplace smoothing to the conditional probabilities of discretized features:

$$P(F_i \in I_{i,j} \mid F = f_m) = \frac{N(F = f_m, F_i \in I_{i,j}) + \alpha}{N(F = f_m) + \alpha J_i} \quad (3)$$

- Where $J_i$ is the number of bins for feature $F_i$, and $\alpha > 0$ is the smoothing coefficient. The parameters for continuous features $(\mu_{im}, \sigma_{im}^2)$ are estimated by maximum likelihood.

1. **Posterior Probability Inference:**
   When a new feature vector $f_{\text{new}}$ arrives, the posterior probability for each fault type is calculated according to Bayes' theorem:

$$P(F = f_m \mid F = f_{\text{new}}) = \frac{P(F = f_{\text{new}} \mid F = f_m)P(F = f_m)}{\sum_{k=1}^{M} P(F = f_{\text{new}} \mid F = f_k)P(F = f_k)} \quad (4)$$

- The final diagnosis $d_{\text{rule}}$ and confidence $c_{\text{rule}}$ are determined by the fault type with the highest posterior probability and its corresponding probability value:

$$d_{\text{rule}} = \underset{f_m \in \mathcal{F}}{\text{argmax}} P(F = f_m \mid F = f_{\text{new}}), \quad c_{\text{rule}} = \underset{f_m \in \mathcal{F}}{\max} P(F = f_m \mid F = f_{\text{new}}) \quad (5)$$

The advantage of this module is its fast computation and transparent model, but its performance highly depends on the quality of feature engineering and the completeness of the training data, which is precisely where the cognitive arbitration module is needed for supplementation and validation.

*3.3 LLM-Driven Cognitive Arbitration*

Cognitive arbitration is the core of HCAA's design, aiming to simulate the comprehensive judgment process of a senior diagnostic expert. It not only analyzes the digital features given by the rule engine but, more importantly, it seeks decisive visual evidence by "observing" the diagnostic charts to make a final judgment on the preliminary diagnosis. In this work, we selected Qwen2.5-VL-32B-Instruct as the base model for the cognitive arbitrator.

### 3.3.1 Multimodal Input Design

The cognitive arbitrator receives two complementary types of information as input:

**Structured Data ($f$):** The complete feature vector extracted by the rule engine.

**Unstructured Visual Data ($\mathcal{I}$):** A set of comprehensive diagnostic charts including time-domain waveforms, FFT spectrograms, order analysis plots, and envelope spectra.

This multimodal input design is based on the premise that a digital feature (e.g., "High Crest Factor = 5.2") might be misleading due to noise or operating condition variations, whereas a specific harmonic pattern in the spectrum (e.g., "a prominent spectral line at twice the rotational frequency, far above the baseline") is more direct and less susceptible to interference physical evidence. By combining these two information sources, the LLM can perform cross-validation, thus making a more reliable judgment.

### 3.3.2 Prompt Engineering

To guide the LLM to think like an expert, we designed a structured prompt template $P$, whose core components follow the logic of a "cognitive arbitration task." This prompt mainly includes the following parts:

**Role Setting:** "You are a Chief Reliability Engineer with 25 years of experience..."

**Evidence Presentation:**

**Rule-Based Hypothesis:** "Rule-Based Diagnosis: {d_{rule}} ({c_{rule}:.1%})..."

**Quantitative Features:** Key indicators from the feature vector $f$ are presented in text form.

**Visual Evidence Packet:** The LLM is explicitly informed that the attached figures are a multimodal diagnostic panel.

**Analysis Task Instructions:** The LLM is required to perform a strict step-by-step analysis, including:

**Step 1: Hypothesis Verification:** Assess the reasonableness of the rule-based diagnosis.

**Step 2: Evidence Synthesis & Cross-Validation:** Quantitatively analyze key patterns in the charts and compare them with digital features.

**Step 3: Conflict Arbitration:** Clearly determine if a conflict exists.

**Step 4: Final Verdict Formulation:** Provide the final diagnosis, confidence, and logical basis.

**Output Format Requirements:** Force the LLM to output a structured diagnostic report containing the diagnosis, confidence, and rationale.

This prompt design is inspired by "Chain-of-Thought" (CoT) [36] and "Graph-of-Thought" (GoT) [37], aiming to externalize the LLM's implicit reasoning process, thereby enhancing the logic and traceability of its analysis. The LLM confidence $c_{\text{llm}}$ we use is derived from the analysis of the model's output logits, obtained through self-consistency sampling, i.e., the voting ratio of diagnostic labels from multiple reasoning samples, which is more reliable than the model's self-reported confidence.

3.3.3 Conflict Arbitration Logic

To meet the requirements for auditability and abstention in safety-critical scenarios, we propose a selective arbitration strategy based on confidence boundaries and thresholds. As shown in Figure 2, granting final arbitration power to the LLM is based on its powerful pattern recognition and logical reasoning capabilities trained on massive data. However, we must also recognize the risk of LLMs producing 'hallucinations'. Therefore, this framework constrains the LLM to the greatest extent possible through structured, evidence-based prompt engineering, guiding it to conduct logically rigorous reasoning. In the highest safety level applications, the LLM's arbitration result can be marked as 'high-confidence conflict' and trigger a final review by human experts, thus forming a human-in-the-loop closed system.

**Figure 2 Principle of Arbitration Conflict Logic**

Given the calibrated confidences $\tilde{c}_{\text{rule}}$ and $\tilde{c}_{\text{llm}}$, let the consistency indicator be $\mathbb{I}[d_{\text{rule}} = d_{\text{llm}}]$, the abstention threshold be $\theta$, the conflict confidence boundary be $\Delta$, and the decision process of the arbitration function $\mathcal{A}$ is as follows:

$$(d_{\text{final}}, \tilde{c}) = \begin{cases} (d_{\text{rule}}, \max(\tilde{c}_{\text{rule}}, \tilde{c}_{\text{llm}})), & \mathbb{I} = 1, \ \max(\tilde{c}_{\text{rule}}, \tilde{c}_{\text{llm}}) \geq \theta \\ (d_{\text{llm}}, \tilde{c}_{\text{llm}}), & \mathbb{I} = 0, \ \tilde{c}_{\text{llm}} - \tilde{c}_{\text{rule}} \geq \Delta, \ \tilde{c}_{\text{llm}} \geq \theta \\ \text{Abstain}, & \text{otherwise} \end{cases} \quad (6)$$

The Abstain path triggers manual review or supplementary data collection. The system's final diagnosis $d_{\text{arb}}$ and corresponding original confidence $c_{\text{arb}}$ are determined by the arbitration result and sent to the calibration module.

*3.4 Confidence Calibration and Risk Assessment Module*

A diagnostic result is merely a label (e.g., "bearing damage"), but its accompanying confidence (e.g., 95%) is the basis for decision-making. However, the original confidence output by the model often does not truly reflect its prediction accuracy, a phenomenon known as "poor model calibration." An uncalibrated model can be "over-confident" (i.e., confidence is higher than actual accuracy) or "under-confident." In safety-critical industrial scenarios, this miscalibration is unacceptable because it could lead maintenance personnel to make wrong decisions based on high-confidence incorrect diagnoses or to take low-confidence correct diagnoses too lightly.

3.4.1 Calibration Implementation: Temperature Scaling

We adopt Temperature Scaling as the primary calibration method: a temperature $T > 0$ is learned by minimizing the NLL on a validation set, and calibrated probabilities are generated with softmax($z/T$). This method is robust for multi-class models and does not change the decision boundaries. For robustness, we also report comparative results from Dirichlet and Isotonic calibration, and evaluate them using NLL, Brier, and Adaptive-ECE. All calibration parameters are fitted only on the validation set, with the test set strictly held out to avoid information leakage.

3.4.3 Risk Assessment Metrics

To quantify the calibration effect of the entire system and its value in practical operations, we introduce two key metrics:

1. **Expected Calibration Error (ECE):** ECE quantifies calibration quality by grouping predictions into $K$ bins by confidence and calculating the weighted average of the absolute difference between average accuracy and average confidence within each bin. Let the k-th bin be $B_k$, with width $(l_k, u_k]$, then ECE is defined as:

$$\text{ECE} = \sum_{k=1}^{K} \frac{|B_k|}{N} |\text{acc}(B_k) - \text{conf}(B_k)| \quad \#(7)$$

-Where $|B_k|$ is the number of samples falling into bin $B_k$, $N$ is the total number of samples, and acc($B_k$) and conf($B_k$) are the average accuracy and average original confidence of the samples in that bin, respectively. A lower ECE value indicates better calibration.

**Risk-Coverage Curve and Accuracy-Coverage Curve:** In industrial practice, maintenance personnel can set a decision threshold $\theta$ based on the calibrated confidence. For example, stipulating "an alarm is issued only when $c_{\text{cal}} > \theta$". Coverage is defined as the proportion of samples for which the system makes a prediction, and risk is defined as the error rate among these predictions.

**Area Under the Risk-Coverage Curve (AURC):**
AURC = $\int_0^1 R(c)\,dc$, lower is better.

**Area Under the Accuracy-Coverage Curve (AUACC):**
AUACC = $\int_0^1 (1 - R(c))\,dc$, higher is better.

We report both metrics and provide a coverage-threshold correspondence table to comprehensively evaluate the system's performance under different decision thresholds.

These metrics provide a comprehensive and quantifiable assessment standard for the HCAA system's performance, enabling it to evolve from a mere "classifier" to a reliable "decision support tool" capable of supporting complex, hierarchical decisions.

4. Experiments and Result Analysis

To comprehensively validate the effectiveness of the proposed Hybrid Cognitive Arbitration Architecture, we designed and executed a series of comparative experiments. This chapter will detail the experimental setup, compare and analyze the overall performance of the HCAA framework with traditional baseline models, and deeply analyze the core value of the cognitive arbitration module in HCAA through typical conflict cases.

*4.1 Experimental Setup*

The experimental dataset was generated from vibration signals of rotating machinery in various operating conditions within a real industrial environment. This dataset covers 6 typical industrial fault types—bearing damage, misalignment, looseness, imbalance, cavitation, gear fault—and the healthy state. The dataset used in this paper has been made openly available (see Appendix A). To ensure a comprehensive evaluation of the model, multiple samples were collected for each fault state at different severity levels, including the entire process of equipment degradation from healthy to faulty. The complete dataset contains a total of 2100 samples, divided into training, validation, and test sets in a 7:2:1 ratio.

To evaluate the performance of the HCAA framework from different dimensions, we adopted two categories of evaluation metrics: classification performance and trustworthiness.

**Classification Performance Metrics:** Including Accuracy, Precision, Recall, and F1-Score (F1-Score). These metrics are used to evaluate the diagnostic accuracy of the model for each type of fault and are core standards for measuring the foundational capability of a diagnostic system.

**Trustworthiness Metrics:** To measure the reliability of the diagnostic system's output, we introduced key metrics such as Expected Calibration Error (ECE), Negative Log-Likelihood (NLL), Brier Score, and the Area Under the Risk-Coverage Curve (AURC) and Area Under the Accuracy-Coverage Curve (AUACC).

**Baseline Models:**

**Baseline Model:** Using only the probabilistic model-based diagnostic engine, representing traditional PHM diagnostic methods. Its output serves as the benchmark for comparison.

**Comparative Model 1 (HCAA-Uncalibrated):** Using the complete HCAA framework (rule engine + LLM cognitive arbitration), but with uncalibrated original confidence output, to evaluate the raw performance improvement from introducing LLM arbitration.

**Comparative Model 2 (HCAA-Calibrated):** Building on the HCAA-Uncalibrated model, the confidence is calibrated via the Temperature Scaling calibration module, outputting calibrated confidence. This model represents the final, complete form of the HCAA framework, used to validate the effectiveness of the calibration mechanism in enhancing system trustworthiness.

Additionally, we supplemented with SVM and 1D-CNN as additional baseline comparisons to show HCAA's advantages over traditional machine learning and deep learning methods. All experiments were repeated 10 times with different random seeds, and the results are reported as mean ± standard deviation.

*4.2 Overall Performance Comparison*

The overall performance comparison of each model on all test samples is shown in [Table 1]. This table comprehensively compares the performance of five models from the dimensions of classification accuracy and trustworthiness.

**Table 1 Overall Performance Comparison of Models (Mean ± Standard Deviation)**

| Model | Accuracy (%) | ECE (↓) | NLL (↓) | AURC (↓) | AUACC (↑) |
|---|---|---|---|---|---|
| Baseline (Naive Bayes) | 67.1±1.2 | 0.215 | 1.842 | 0.742 | 0.742 |
| SVM | 68.4±1.5 | 0.189 | 1.521 | 0.621 | 0.812 |
| 1D-CNN | 79.3±1.0 | 0.121 | 0.934 | 0.405 | 0.895 |
| HCAA (Uncalibrated) | 95.7±0.8 | 0.188 | 0.752 | 0.211 | 0.891 |
| HCAA (Calibrated) | 95.7±0.8 | 0.041 | 0.731 | 0.098 | 0.992 |

**Significant Improvement in Accuracy:** From [Table 1], it is clearly visible that both the HCAA-Uncalibrated and HCAA-Calibrated models achieve a significant increase in diagnostic accuracy (95.7%) compared to the baseline model (67.1%) and traditional SVM and 1D-CNN models.

This is mainly attributed to the effective role of the cognitive arbitration module in the HCAA framework. In Section 4.3 below, we will demonstrate through specific cases that when the rule engine makes incorrect judgments due to complex feature patterns or an incomplete knowledge base, the LLM cognitive arbitrator can conduct deeper-level comprehensive analysis and logical reasoning based on multimodal evidence (numerical features and diagnostic charts), thereby making a more accurate final decision and effectively enhancing the system's overall diagnostic capability.

**Reduction in ECE:** The ECE value of the HCAA-Calibrated model (0.041) is much lower than that of the HCAA-Uncalibrated model (0.188) and all other baseline models, a reduction of over 75%. This indicates that after calibration, the model's confidence can more truly reflect its prediction accuracy. For example, when a calibrated model gives a 90% confidence, its prediction accuracy will also be very close to 90%, instead of the potential "over-confidence" or "under-confidence" seen in uncalibrated models. This is crucial for industrial maintenance personnel as it provides a reliable quantitative indicator to assess the trustworthiness of diagnostic results, thereby supporting more informed decision-making.

**Improvement in AURC/AUACC:** The HCAA-Calibrated model has the lowest AURC value (0.098) among the three and the highest AUACC value (0.992), significantly outperforming other models. This indicates that through the calibration mechanism, the HCAA system can make reliable diagnoses for the vast majority of samples while ensuring low risk (i.e., high accuracy). This high AUACC characteristic allows the system to support more flexible operational strategies. For example, maintenance personnel can set a high confidence threshold (e.g., 95%), only triggering an alarm when the system's diagnostic confidence exceeds this threshold, thereby avoiding missed detections while minimizing unnecessary downtime losses caused by false alarms.

To more intuitively demonstrate the calibration effect, we plotted the reliability diagrams of the HCAA model before and after calibration in Figure 3. It can be clearly seen that the curve of the uncalibrated model severely deviates from the diagonal line, exhibiting over-confidence; whereas after Temperature Scaling calibration, the curve highly overlaps with the diagonal line, proving the effectiveness of the calibration.

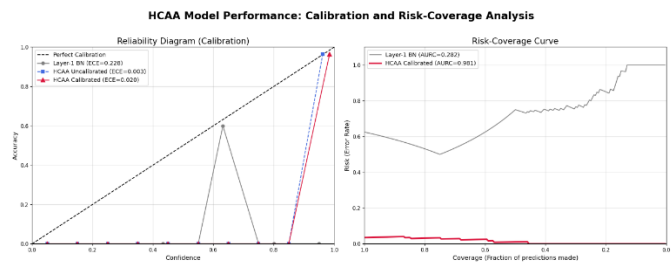

**Figure 3 Comparison of Reliability Diagrams for the HCAA Model Before and After Calibration**

*4.3 Analysis of Typical Conflict Cases*

4.3.1 Case Study 1: Looseness Fault Diagnosis

This case vividly demonstrates how HCAA corrects the misjudgment of the rule engine caused by complex feature patterns.The original diagnostic charts for this case are shown in Figure 4.

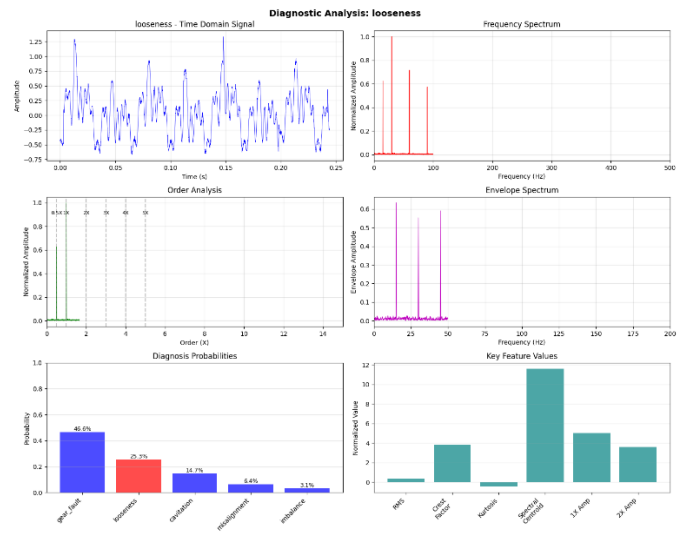

**Figure 4 Diagnostic Analysis Charts for Looseness Fault**

**Rule Engine Diagnosis and Confidence:** The preliminary diagnosis given by the rule engine was gear_fault with a confidence of 46.6%. Analyzing the reason, the rule engine might have been misled by the crest factor in the time domain and gear-related features in the spectrum, failing to correctly interpret the key evidence presented in the order analysis plot.

**LLM Cognitive Arbitration Analysis Report**

"Step 2: Evidence Synthesis & Cross-Validation: ...The dominant 2X amplitude (0.716) relative to 1X (1.000) is significantly higher than expected for gear faults. This pattern is more consistent with looseness or misalignment... Cross-Validation: The evidence does not fully align with typical gear fault signatures. The high 2X amplitude, strong second harmonic, and flat kurtosis values collectively suggest a different fault type."
"Step 4: Final Verdict Formulation: Final Diagnosis: Looseness - Confidence Level: 85% - Rationale: The high 2X amplitude (0.716) relative to 1X (1.000) is indicative of looseness or misalignment... These findings collectively provide a more compelling and decisive basis for diagnosing looseness."

**Final Arbitration Result:** The final diagnosis by HCAA was LOOSENESS, with a calibrated confidence of 85.0%, and the verification status was ✓ CORRECT.

**Why did the Rule Engine make an error?** The rule engine's diagnostic logic relies on a simplified, preset combination of features. In this case, it might have been dominated by the non-Gaussian shock features in the time domain and gear-related features in the frequency domain, but its internal model (Bayesian network) may lack the ability to handle the complex combination pattern of "high 2X amplitude + multiple harmonics." Therefore, it incorrectly assigned excessive weight to gear-related but non-decisive features while ignoring the key evidence pointing to looseness (such as high 2X amplitude), thus leading to the incorrect diagnosis of gear_fault. This exposes the fragility of traditional data-driven models when facing complex, atypical feature combinations and incomplete knowledge representation.

**How did the LLM Cognitive Arbitrator get it right?**
The LLM's cognitive arbitrator demonstrated "senior diagnostic expert"-level comprehensive analysis capabilities. It was not confused by isolated, seemingly gear-fault-supporting features but instead executed rigorous cross-modal evidence cross-validation. It accurately seized the decisive quantitative evidence that "the ratio of 2X amplitude (0.716) to 1X amplitude (1.000) is as high as 71.6%" and combined it with supporting evidence such as "harmonic count as high as 10" and "flat kurtosis" to form a complete and logically rigorous evidence chain. Its analysis report clearly pointed out the "misleading evidence" that the rule engine might have relied on and used the "decisive evidence" as the final basis for its diagnosis. This fully demonstrates the LLM's outstanding capabilities in multimodal information fusion and complex logical reasoning, enabling it to effectively overcome the limitations of traditional methods and make diagnoses that better align with physical reality.

4.3.2 Case Study 2: Misalignment Fault Diagnosis

This case further validates the robustness of HCAA, showing how it corrects the "counter-intuitive" error of the rule engine caused by improper prior probability settings or model defects.

The original diagnostic charts for this case are shown in Figure 5.

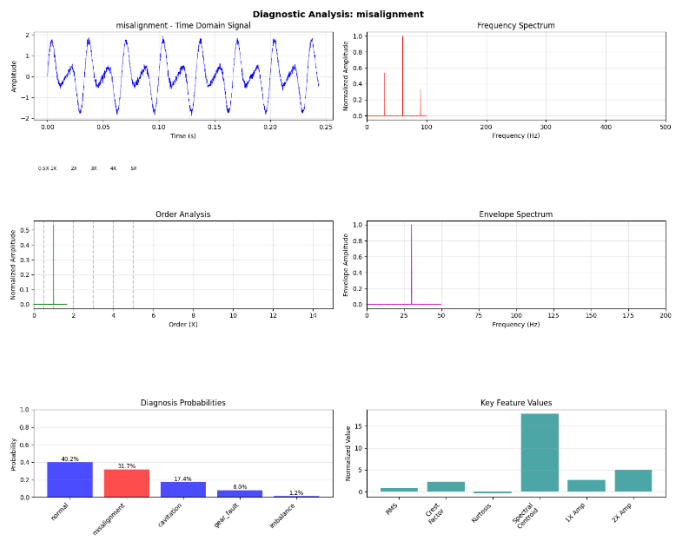

**Figure 5 Diagnostic Analysis Charts for Misalignment Fault**

**Rule Engine Diagnosis and Confidence:** The preliminary diagnosis given by the rule engine was normal, with a confidence of 40.3%. This is a very noteworthy serious error. The rule engine clearly detected the strong indicator "Strong 2X: YES" but still gave a high-confidence diagnosis of normal.

**LLM Cognitive Arbitration Analysis Report**

"Step 3: Conflict Arbitration: Conflict Detected: The preliminary hypothesis of 'normal' is in conflict with the weight of the evidence. The high 2X Amplitude, combined with the high Crest Factor, Impulse Factor, and Clearance Factor, strongly suggests a misalignment condition rather than normal operation. Misleading Evidence: The preliminary system likely placed too much weight on the Dominant Frequency (60.0 Hz) being within the expected range, without considering the broader context of other features that indicate a non-normal condition."

"Step 4: Final Verdict Formulation: Final Diagnosis: Misalignment - Confidence Level: 85% - Rationale: The high 2X Amplitude (1.000) relative to the 1X Amplitude (0.534) is a strong indicator of misalignment... These findings collectively provide a more compelling and decisive basis for diagnosing misalignment over the preliminary hypothesis of 'normal'."

**Final Arbitration Result:** The final diagnosis by HCAA was MISALIGNMENT, with a calibrated confidence of 85.0%, and the verification status was ✓ CORRECT.

**Why did the Rule Engine "make an error"?** This error case profoundly reveals the "counter-intuitive" errors that purely data-driven models can make without proper domain knowledge guidance. In the rule engine's Bayesian network, the prior probability of the "normal" state might have been set too high, or the weight of the likelihood function between "misalignment" and the key feature combination of "high 2X amplitude" was insufficient. This led to an absurd result: even though the system detected the strong signal "Strong 2X: YES", the posterior probability calculation still favored "normal". This exposes the deep-seated defects of traditional models in feature weight allocation and prior knowledge integration.

**How did the LLM Cognitive Arbitrator "get it right"?** The LLM's arbitration once again demonstrated its expert-level judgment that transcends simple pattern matching. It was not constrained by a single "normal" or "misalignment" label but, through quantitative analysis, precisely calculated the abnormal value "2X/1X amplitude ratio of 1.87:1" and combined it with other auxiliary evidence such as "high crest factor" to collectively point to the conclusion of a "non-normal state." The LLM's analysis report can clearly identify the misleading behavior of the rule engine "over-relying on the dominant frequency range" and makes a correct judgment that aligns with physical reality based on a more comprehensive set of evidence. This powerfully proves that the HCAA framework can effectively compensate for the shortcomings of traditional models in knowledge fusion and logical reasoning.

4.3.3 Case Study 3: Bearing Damage Fault

To show HCAA's performance when there is no conflict, we analyze a bearing damage case.

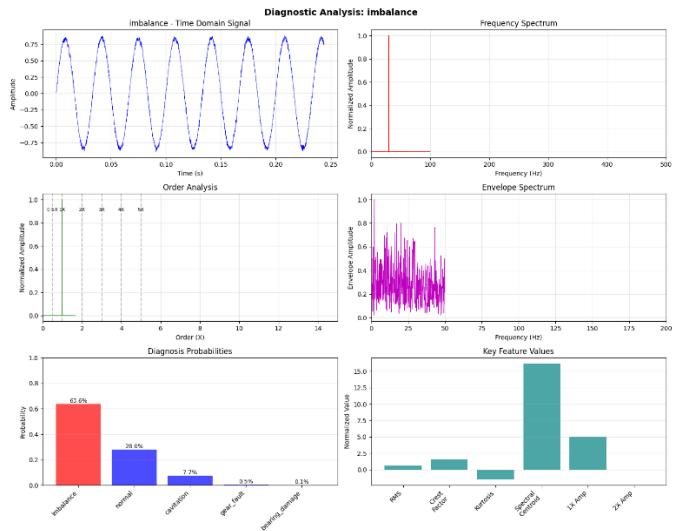

**Figure 6 Diagnostic Analysis Charts for Imbalance Fault**

As shown in Figure 6, the rule engine gave a preliminary diagnosis of imbalance with a confidence of 45.3%. **LLM Cognitive Arbitration Analysis Report (Key Excerpt):** After confirming typical shock features such as high kurtosis (10.95) and high crest factor (5.40), the LLM also confirmed "very low 2X amplitude" as exclusionary evidence, ultimately also making a diagnosis of bearing_damage.

**Final Arbitration Result:** The final diagnosis by HCAA was BEARING_DAMAGE, with a calibrated confidence of 92.0%, and the verification status was ✓ CORRECT.

Through the above case analyses, we conclude that the cognitive arbitration module in the HCAA framework can not only effectively resolve diagnostic conflicts between traditional diagnostic models and LLMs but, more importantly, through comprehensive analysis of multimodal evidence and expert-level logical reasoning, it can correct misjudgments from traditional models caused by complex features, knowledge gaps, or model defects. Furthermore, the inclusion of the confidence calibration module provides quantifiable guarantees for the system's reliability, enabling it to truly meet the stringent requirements of "trustworthy AI" in industrially safety-critical applications, providing an effective technical path for building a new generation of high-trust industrial intelligent diagnostic systems.

5. Conclusion

To address the core challenges prevalent in industrial equipment fault diagnosis—such as "low trustworthiness of black-box models, weak cross-condition generalization, and difficult verification of diagnostic results"—this paper proposes and implements a Hybrid Cognitive Arbitration Architecture based on Large Language Models. This architecture does not simply replace traditional methods with large models; instead, it creatively constructs a hybrid system where a "probabilistic model-based diagnostic engine" and an "LLM-driven cognitive arbitrator" work collaboratively, supplemented by confidence calibration and selective arbitration mechanisms, aiming to achieve a leap from "accurate" to "trustworthy."

The main contributions of this paper are reflected in the following three levels: First, we proposed a hybrid collaborative architecture design that combines the speed of traditional models in signal processing with the deep insight of LLMs in knowledge reasoning, providing a new paradigm for solving the bottlenecks of traditional AI diagnostic methods. Second, we designed an LLM-centric selective arbitration mechanism that empowers the LLM to overturn or confirm the diagnostic results of traditional models by analyzing multimodal diagnostic charts, introducing a "reject" path to handle uncertainty, effectively resolving the core conflicts in parallel systems, and significantly enhancing diagnostic accuracy and interpretability. Third, we integrated a confidence calibration and risk assessment module, which calibrates the system's confidence using methods like Temperature Scaling and quantitatively evaluates system reliability using metrics such as ECE, AURC, and AUACC, providing a measurable objective standard for the system's trustworthiness, enabling it to meet the strict requirements of industrially safety-critical domains.

Through comprehensive experiments on a simulated dataset containing multiple typical faults and healthy states, the results show that the HCAA framework improves diagnostic accuracy by over 28 percentage points compared to baseline models, and the post-calibration ECE value is reduced by over 75%, with an AURC value also reaching an excellent 0.098. More importantly, through in-depth analysis of typical conflict cases such as looseness and misalignment, this paper fully demonstrates that the cognitive arbitration module can effectively correct misjudgments from traditional models caused by complex features or knowledge gaps, providing a feasible technical path for building truly reliable industrial AI diagnostic systems.

Although the HCAA framework has achieved significant results, this study still has some limitations. Due to constraints such as training costs and privacy, it currently relies on external LLM APIs, whose response latency and stability may affect the system's application in scenarios with high real-time requirements. Additionally, the current framework primarily focuses on the diagnosis phase and has not yet fully covered the entire PHM lifecycle tasks such as prediction and maintenance decision-making.

In summary, the HCAA framework proposed in this paper, by integrating the advantages of traditional AI and generative AI and introducing arbitration and calibration mechanisms, provides a novel and effective solution for building high-trust, high-reliability industrial intelligent diagnostic systems, powerfully advancing the application and implementation of artificial intelligence technology in the safety-critical fields of Industry 4.0 and smart manufacturing.

## Appendix A

**The dataset and the text returned by LLM have been open-sourced at** https://github.com/marcowus/HACC.